\documentclass[showpacs,floatfix]{revtex4-1}
\usepackage{graphicx,psfrag,amsmath,amssymb,amsfonts,latexsym,color,dcolumn,graphpap}

\begin{document}

\title{Decoherence of a solid-state qubit by different noise correlation spectra}

\author{Paula I. Villar\footnote{paula@df.uba.ar}
and Fernando C. Lombardo\footnote{lombardo@df.uba.ar}}
\affiliation{Departamento de F\'\i sica {\it Juan Jos\'e
Giambiagi}, FCEyN UBA and IFIBA CONICET-UBA, Facultad de Ciencias Exactas y Naturales,
Ciudad Universitaria, Pabell\' on I, 1428 Buenos Aires, Argentina}

\date{today}

\begin{abstract}
The interaction between solid-state qubits and their environmental degrees of freedom  produces 
non-unitary effects like decoherence and dissipation. Uncontrolled
decoherence is  one of the main obstacles that must be overcome in quantum information 
processing. We study the dynamically decay of coherences in a solid-state qubit by means of the use of a  
master equation.  We analyse the effects induced
by thermal Ohmic environments and low-frequency 1/f noise. 
We focus on the effect of longitudinal and transversal noise on the superconducting qubit's
dynamics. Our results can be used to design experimental future setups
when manipulating superconducting qubits.
\end{abstract}

\pacs{03.65.Yz, 03.67.Hk, 75.10.Jm, 74.50.+r}

\maketitle

\newcommand{\beq}{\begin{equation}}
\newcommand{\eeq}{\end{equation}}
\newcommand{\dalam}{\nabla^2-\partial_t^2}
\newcommand{\mbf}{\mathbf}
\newcommand{\itm}{\mathit}
\newcommand{\beqa}{\begin{eqnarray}}
\newcommand{\eeqa}{\end{eqnarray}}

\section{Introduction}
\label{intro}

The scaling-down of microelectronics into the nanometer range will inevitably make
quantum effects such as tunnelling and wave propagation important. The use of these
quantum devices in gate operations enhances the need of controlling decoherence.
Noise from the environment may cause fluctuations in both qubit amplitude and phase,
leading to relaxation and decoherence. 
External perturbations can influence a two-level system in typically two ways: 
either shifting the individual energy levels (which changes the transition energy and therefore,
the qubit's phase) or inducing energy levels transitions  (which changes the
level populations). 
Decoherence is a major hurdle in realising scalable quantum technologies in the solid-state. 

Decoherence in qubit systems falls into two general categories. 
One is an intrinsic decoherence caused by constituents in the qubit system, 
and the other is an extrinsic decoherence caused by the interaction with uncontrolled
degrees of freedom, such as an environment. Understanding the mechanisms of decoherence and
achieving long decoherence times is crucial for many fields of science and applications 
including quantum computation and quantum information \cite{nielsen}.
Most theoretical investigations of how the system is affected
by the presence of an environment have been done using
a thermal reservoir, usually assuming Markovian statistical
properties and defining bath correlations \cite{paz,weiss}. 
However, there has been some growing interest
in modelling more realistic environments, sometimes called
composite environments, or environments out of thermal equilibrium \cite{pra05,pra83,pra13}.

Lately, there have been many studies focusing in decoherence in a solid-state qubit. 
The same physical structures that make these superconducting qubits easy to manipulate, measure, 
and scale are also responsible for coupling the qubit to other electromagnetic degrees of freedom that can 
be source of decoherence via noise and dissipation. Thus, a detailed mechanism of decoherence and
noise due to the coupling of Josephson devices to external noise sources is still required.
In Ref.\cite{ithier} there is a detailed experimental study, inspired from NMR, in order to characterise decoherence in a 
particular superconducting qubit, known as quantronium. An analysis of different noise sources is presented on the 
basis of the study of the environmental spectral densities.
In \cite{paladino} authors reviewed  the effect of 1/f noise in nano-devices with emphasis
on implications for solid-state quantum information. 
It has been shown that low frequency noise is an important source of decoherence for superconducting qubits.
In this detailed review of experimental solid state quantum information, authors dealt with the decoherence effect of $1/f$ noise
through a purely dephasing phenomenological approach.
In Ref.\cite{nesterov}, the influence of $1/f$ noise by random dephasing telegraph processes was also modelled, 
showing that depending
on the parameters of the environment, the model can describe both Gaussian and non-Gaussian effects
of noise. Ref.\cite{zhou}  presented a phenomenological model for superconducting qubits subject
to noise produced by two-state fluctuators whose coupling to the qubit are all roughly the same.
In all cases mentioned, authors analysed the qubit's evolution under pure dephasing conditions, since
it is the case where the dynamics can be analytically obtained.
 When we departure from pure dephasing, i.e., when the noise-system interaction is not longitudinal, 
 no exact analytic solution for the system time
evolution is available even for the spin-fluctuator model. 
In \cite{pekola}, authors resorted to a master equation approach to study 
the influence of an environment on a two-level system with an adiabatically changing
external field where they have also considered temperature effects.
However, authors resorted to a secular approximation
for the ground state only, assuming that the excitation rates are exponentially
small with respect to the 
relaxation ones in the quasi-stationary regime. They obtain that in the zero-T limit,
the ground state of the system is not affected by the environment and hence, there is no
decoherence in that case. This fact can not be true since
decoherence at zero temperature does occur.
There are simple examples in Literature which demonstrate that decoherence is induced even by a 
reservoir  at zero temperature \cite{pla,pla2,pre,jppdromero}. In general, a
small system coupled to an environment fluctuates even in the
zero-T limit. These fluctuations can take place without generating 
an energy trace in the bath. The fluctuations in energy of
the small system are a peculiar fact of the entanglement with
the quantum environment. 
 However, the suppression
of the interferences is not as fast as it is at high temperature
limit. In the latter case, it is expected to happen, for a quantum Brownian particle of mass $M$,  
 at times of
$O(1/2M \gamma_0 k_B T L)$ 
 while it occurs at times bigger
 than $O(1/\gamma_0)$  when the environment is at zero temperature \cite{pla} (where $\gamma_0$ 
is the dissipation constant, $T$ is the environmental temperature, and $L$ the distance between classical 
trajectories of the particle). 
 In \cite{martinis} authors study decoherence effects on a superconducting 
qubit due to bias noise making use of a semiclassical noise model that includes the effect 
of zero-point and thermal effects. 

In the review article by Makhlin et al \cite{makhlin}, a dephasing model is considered 
in which the spin-system is coupled to an Ohmic environment at thermal equilibrium. They show that 
the dephasing time is shorter than the measurement time, and they estimate the mixing time, i. e. 
the time scale on which transitions induced by the measurement occur. They analyse the dephasing rate 
in an exactly solvable limit of a qubit with no tunnelling term. 

In \cite{berger}, noise is described by fluctuations in the effective magnetic field which are directed
either in the $z$ axis -longitudinal noise- or in a transverse direction -transversal noise.
Both types of noise have been phenomenologically modelled by making different assumptions on these fluctuations,
such as being due to a stationary, Gaussian and Markovian process. The results obtained were applied on adiabatic geometric
phase. They experimentally study which noise -either radial or angular- contribute to the geometric (Berry)
and dynamical phase, observing that the radial noise is more destructive.

In this context, we aim to gather the main of these features (and the ones that have been neglected by the sake of simplicity) 
that have been studied separately in a single approach
by the study of decoherence process from a master equation for a superconducting qubit. 
By considering a general approach of the qubit and environment, we include different channels of decoherence
(purely dephasing and decoherence by anomalous diffusion coefficients).
This approach does not admit an analytical solution in the end but by means of a numerical solution of
the reduced density matrix we can obtain  the dynamics
of the qubit. This allows to focus on different noise sources 
and different couplings to the environment and gives a common framework to
study the decoherence process of the qubit. This is particularly important 
in the case of very low temperature fluctuations,
which sometimes are considered to be harmless to the unitary evolution of the qubit.
In such a complete case, the evolution of the system undergoes dephasing and relaxation
leading to serious problems in maintaining the quantum state.

In this article, we shall present a fully quantum
open-system approach to analyse the non-unitary dynamics
of the solid-state qubit when it evolves under
the influence of external fluctuations. We consider the qubit
coupled in longitudinal and transversal directions, that is a model where populations and coherences 
are modified by the environment. We study the dynamics and decoherence-induced
process on the superconducting qubit. 
The paper is organised as follows: In Sec.  \ref{model},
we develop a general quantum open-system model in order
to consider different types of fluctuations (longitudinal and/or
transverse) that induce decoherence on the main system. 
In Sec. \ref{master} we present the master equation approach
for a general coupling of a superconducting, and 
numerically compute the non-unitary evolution
characterised by fluctuations, dissipation, and decoherence.
This gives us a complete insight into the state of the system:
complete knowledge of different dynamical time scales and
analysis of the effective role of noise sources inducing
diffusion and dissipation.
 We also analyse low-frequency $1/f$ noise as coming from a fluctuator environment, by defining the
corresponding spectral density. The comprehension of the
decoherence and dissipative processes should allow their further suppression in future qubits designs or 
experimental setups.
 In Sec. \ref{ohmic},  we analyse the effect induced in the system 
by thermal Ohmic environments, through a non-purely dephasing process, including the case 
of a zero-temperature bath. In Sec.\ref{1f}
we study the decoherence induced by a $1/f$ noise.  In both cases, we 
particularly study the difference between the longitudinal and transversal couplings and provide
analytical estimations of decoherence time when possible. 
Finally in Sec. \ref{conclusions}, we summarise
our final remarks.

%************************************************************************************
%************************************************************************************
%************************************************************************************
%************************************************************************************

\section{Model for a solid-state qubit}
\label{model}

Superconducting qubits are made of inductors, capacitors, 
and Josephson junctions (JJ) \cite{Josephson}, where a JJ 
consists of a thin layer of insulator between superconducting
electrodes. A quantum circuit consisting only of inductors 
and capacitors gives rise to parabolic energy potentials 
exhibiting equally spaced energy levels, which are
not practical for qubits. The JJ provides the necessary
nonlinearity to the system, leading to non-parabolic energy 
potentials with unequally spaced energy levels such that 
two out of several energy levels, serving as the qubit
states  $|0\rangle$ and $|1\rangle$, can be isolated.
Experimental observation of Rabi oscillations in driven quantum circuits
have shown several periods of coherent oscillations, confirming
the validity of the two-level approximation and possibility of coherently
superimpose the computational two states of the system. Nevertheless, the
unavoidable coupling to a dissipative environment surrounding the circuit 
represents a source of relaxation and decoherence that limits the performances of
the qubit for quantum computation tasks. Therefore, for the implementation
of superconducting circuits as quantum bits, it is necessary to understand the
way the system interacts with the environmental degrees of freedom, and to
reduce their effect, if possible.

When the two lowest energy levels of a current biased Josephson
junction are used as a qubit, the qubit state can be fully manipulated with low and microwave 
frequency control currents. Circuits presently 
being explored combine in variable ratios the Josephson effect and single Cooper-pair charging effects.
In all cases the Hamiltonian of the system can be written as,
$
H = \frac{\hbar}{2} \omega_a \sigma_z + \hbar \Omega_{\rm R} \cos(\omega t + \varphi_{\rm R}) \sigma_x$, 
where $\hbar \Omega_{\rm R}$ is the dipole interaction amplitude between the qubit and the microwave 
field of frequency $\omega$ and phase $\varphi_{\rm R}$. $\Omega_{\rm R}/2\pi$ is the Rabi frequency. 
This Hamiltonian can be transformed to a rotating frame at the frequency $\omega$ by means of an
unitary transformation defined by the operator $U = \exp{(i \omega t\sigma_z/2)}$ \cite{leek} and, after 
the rotating wave approximation (ignoring terms oscillating at $2\omega$), resulting in a new effective 
Hamiltonian of the form
\begin{equation}
H_{\text{eff}} = \frac{\hbar}{2} \left(\Delta \sigma_z + \Omega_x \sigma_x + \Omega_y \sigma_y \right) ,
\label{Hqubitdiag}
\end{equation}
where $\Omega_x = \Omega_{\rm R} \cos\varphi_{\rm R}$ and  $\Omega_y = \Omega_{\rm R} \sin\varphi_{\rm R}$. 
Then, we shall consider the dynamics of a generic two-level system steered by a system's Hamiltonian of the 
type (where we have set $\hbar = 1$ all along the paper)
\begin{equation}
H_{\text{Total}} = H_{\text{eff}}  + H_{\rm int} + H_{\cal E},\label{HT}\end{equation}
where $H_{\cal E}$ is the Hamiltonian of the bath. The interaction
Hamiltonian (in the rotating frame) is thought as some longitudinal and transverse noise coupled to the main system:
\begin{equation}
H_{\rm int} = \frac{1}{2} \left( {\hat {\delta\omega_1}} \sigma_x +  
{\hat {\delta\omega_2}} \sigma_y + {\hat {\delta\omega_0}} \sigma_z\right).
\label{Hint}
\end{equation}

By considering this interaction Hamiltonian we are implying that the superconducting qubit is coupled to
the environment by a coupling constant in the $\hat{z}$ direction, called longitudinal direction, affecting only the coherences; 
and a coupling in the 
transverse directions ($\hat{x}$ and $\hat{y}$), which modify populations and coherences in a different rate. This type of coupling is a generalisation of the bidirectional 
coupling recently used in \cite{pra89}
to compute the geometric phase of a superconducting qubit. 
It is important to note that the derivation of  a master
equation has not been done before for a solid-state qubit.

%&&&&&&&&&&&&&&&&&&&&&&&&&&&&&&&&&&&&&&&&&&&&&&&&&&&&&&&&&&&&&&&&&&&&&&&&&&&&&&&&&&&&&&&&&&&&&&
%&&&&&&&&&&&&&&&&&&&&&&&&&&&&&&&&&&&&&&&&&&&&&&&&&&&&&&&&&&&&&&&&&&&&&&&&&&&&&&&&&&&&&&&&&&&
\section{Master equation approach}
\label{master}

We derive a master equation for general noise terms 
${\hat {\delta\omega_1}}$, ${\hat {\delta\omega_2}}$ and ${\hat {\delta\omega_0}}$. 
We consider a weak coupling 
between system and environment and the bath sufficiently large to stay in a  
stationary state. In other words, the total state $\rho_{\cal SE}$ (system and environment) 
can be split as 
$\rho_{\cal SE} \approx \rho(t) \times \rho_{\cal E}$,
for all times. It is important to stress that due to the Markov regime, we restrict to cases for which 
the self-correlation functions generated at the environment (due to the coupling interaction) would 
decay faster than typical variation scales in the system \cite{breuer}. In the interaction picture, 
the evolution of the total state 
is ruled by the Liouville equation
\begin{equation}
{\dot \rho}_{\cal SE} = -i \left[ H_{\rm int}, \rho_{\cal SE} \right],
\end{equation} where we have denoted the state $\rho_{\cal SE} $ in the interaction picture in the same 
way than before, just in order to simplify notation. 
A formal solution of the Liouville equation can be obtained perturbatively using the Dyson expansion.

From this expansion, one can obtain a perturbative master equation, up to second order in 
the coupling constant between system and 
environment for the reduced density matrix $\rho = {\mbox Tr}_{\cal E} \rho_{\cal SE}$. In the 
interaction picture the formal solution reads as
\begin{equation}
\rho (t) \approx  \rho (0) - i \int_0^t ds {\mbox Tr}_{\cal E} \left(\left[H_{\rm int}(s), 
\rho_{\cal SE}  (0)\right]\right) -\int_0^{t}ds_1 \int_0^{s_1} ds_2  {\mbox Tr}_{\cal E} \left(\left[H_{\rm int}(s), 
\left[H_{\rm int}(t), \rho_{\cal SE} (0)\right]\right]\right).
\nonumber \end{equation}

In order to obtain the full master equation for the qubit, it is necessary to perform the temporal 
derivative of the previous equation and assume that the system and the environment are not initially correlated.
In addition, we consider that the ${\hat {\delta\omega_i}}$ of the $H_{\rm int}$ (Eq.(\ref{Hint}))
are operators acting only on the Hilbert space of the environment
(and the Pauli matrices applied on the system Hilbert space). Finally, the master equation explicitly reads, 

\begin{eqnarray}
\dot\rho &=& -i \left[H_{\rm eff}, \rho\right] - d_{xx}(t) \left[\sigma_x,\left[\sigma_x,\rho\right]\right] 
- f_{xy}(t) \left[\sigma_x,\left[\sigma_y,\rho\right]\right] 
- f_{xz}(t) 
\left[\sigma_x,\left[\sigma_z,\rho\right]\right] \nonumber \\ 
&-&  f_{zx}(t) \left[\sigma_z,\left[\sigma_x,\rho\right]\right] 
-  f_{yx}(t) \left[\sigma_y,\left[\sigma_x,\rho\right]\right]  
- d_{yy}(t) \left[\sigma_y,\left[\sigma_y,\rho\right]\right]
- f_{yz}(t) \left[\sigma_y,\left[\sigma_z,\rho\right]\right] \label{mastereq} \\
&-& f_{zy}(t) \left[\sigma_z,\left[\sigma_y,\rho\right]\right]
- d_{zz}(t) \left[\sigma_z,\left[\sigma_z,\rho\right]\right] 
+  i \gamma_{xy} \left[\sigma_x,\left\{\sigma_y,\rho\right\}\right]  
+ i \gamma_{xz} \left[\sigma_x,\left\{\sigma_z,\rho\right\}\right] 
+ i \gamma_{zx} \left[\sigma_z,\left\{\sigma_x,\rho\right\}\right]  \nonumber \\ &+& i \gamma_{zy} \left[\sigma_z,\left\{\sigma_y,\rho\right\}\right]
+  i \gamma_{yx} \left[\sigma_y,\left\{\sigma_x,\rho\right\}\right]  
+ i \gamma_{yz} \left[\sigma_y,\left\{\sigma_z,\rho\right\}\right], \nonumber
\end{eqnarray}
where the noise effects are included as the normal ($d_{xx}$, $d_{yy}$ and $d_{zz}$) and anomalous diffusion 
coefficients ($f_{ij}$ with $i,j=x,y,z$). The dissipative effects are considered in the corresponding coefficients 
($\gamma_{ij}$ with $i,j=x,y,z$). Thus, Eq.(\ref{mastereq}) considers
both diffusion and dissipation effects for a superconducting qubit with couplings as in Eq.(\ref{Hint}),
\begin{widetext}
\begin{eqnarray}
d_{xx}(t) &=& \int_0^t ds\,\nu_1(s)\,X_1(-s),~~~d_{yy}(t)=\int_0^t ds\,\nu_2(s)\,Y_2(-s),~~~
d_{zz}(t) =\int_0^t ds\,\nu_0(s)\,Z_0(-s),\nonumber \\
f_{xy}(t) &=& \int_0^t ds\,\nu_1(s)\,Y_1(-s), ~~~
f_{xz}(t) = \int_0^t ds\,\nu_1(s)\, Z_1(-s),~~~
f_{zx}(t) = \int_0^t ds\,\nu_0(s)\, X_0(-s),\nonumber \\
f_{zy}(t) &=&\int_0^t ds\,\nu_0(s)\,Y_0(-s),~~~
f_{yx}(t) = \int_0^t ds\,\nu_2(s)\, X_2(-s),~~~
f_{yz}(t) = \int_0^t ds\,\nu_2(s)\, Z_2(-s) \label{coeficientes}  \\
\gamma_{xy}(t) &=& \int_0^t ds \, \eta_1(s)  \, Y_1(-s),~~~
\gamma_{xz}(t) = \int_0^t ds \, \eta_1(s)  \, Z_1(-s),~~~ 
\gamma_{yx}(t) = \int_0^t ds \, \eta_2(s)  \, X_2(-s),  \nonumber \\
\gamma_{zx}(t) &=& \int_0^t ds \, \eta_0(s)  \, X_0(-s),
~~~ \gamma_{yz}(t) = \int_0^t ds \, \eta_2(s)  \, Z_2(-s),~~~{\mathrm{and}}~~~
\gamma_{zy}(t) = \int_0^t ds \, \eta_0(s)  \, Y_0(-s) \nonumber.
 \end{eqnarray}
\end{widetext}

These coefficients are defined in terms of  
the noise and dissipation kernels, $\nu(t)$ and $\eta(t)$, respectively.
These kernels are generally defined, for unspecified operators ${\hat {\delta\omega_0}}(t)$,
${\hat {\delta\omega_1}}(t)$
and ${\hat {\delta\omega_2}}(t)$, as
\begin{equation}
\nu_{a}(t) = \frac{1}{2} \langle\left\{{\hat {\delta\omega_{a}}}(t),
{\hat {\delta\omega_{a}}}(0)\right\}\rangle ,\label{noise}
\end{equation}
\begin{equation}
\eta_{a}(t) =   \frac{1}{2}  \langle\left[ {\hat {\delta\omega_{a}}}(t),
{\hat {\delta\omega_{a}}}(0)\right]\rangle, \label{dissipation}
\end{equation}
with $a=1,2,0$.
The functions $X_{a}, Y_{a}$, and $Z_{a}$ appearing in Eq.(\ref{coeficientes}) 
are given in the Appendix.
 It is easy to check that if the Rabi frequency is zero and $\delta \hat \omega_1 =0=\delta \hat \omega_2$, 
we recover the dynamics of a spin-$1/2$ precessing a biased field Bloch vector ${\bf R}$. \\

The intention is to study environmental-induced decoherence by means of the use of the master equation. 
We will mainly concentrate on two types of external noise sources. On one side, we shall consider that the environment 
is characterised by an Ohmic spectral density, 
as the one commonly used in models of Quantum Brownian Motion (QBM) 
or in the well known spin-boson model \cite{leggett, pra}. 
In these examples, the environment is represented by an infinite set of harmonic oscillators at thermal equilibrium.
On the other side, we shall analyse decoherence induced effects coming from spin-environments, 
for example spin-fluctuator models, that give us the possibility to study $1/f$ noise-effects 
via the master equation approach, without resorting to classical statistical 
evolutions or phenomenological models. Once the coefficients in 
Eqs.(\ref{coeficientes}) are defined by the use of the corresponding spectral correlations Eqs.(\ref{noise}) and (\ref{dissipation}), 
we can numerically solve the master equation and obtain the evolution 
in time of the reduced density matrix. \\

%%%%%%%%%%%%%%%%%%%%%%%%%%%%%%%%%%%%%%%%%%%%%%%%%%%%%%%%%%%%%%%%%%%%%%%%%%%%%
%%%%%%%%%%%%%%%%%%%%%%%%%%%%%%%%%%%%%%%%%%%%%%%%%%%%%%%%%%%%%%%%%%%%%%%%%%%%%%%

\section{ Ohmic environment} 
\label{ohmic}

A relevant contribution to decoherence in solid-state qubits, is introduced by 
the electromagnetic noise of the control circuit, typically Ohmic noise at low frequencies. 
In this Section, 
we model this kind of environments by means of an infinite set of harmonic oscillators with 
an Ohmic spectral density. An environment composed by harmonic oscillators at thermal 
equilibrium at temperature T is 
commonly introduced in order to take into account dissipative effects, 
additionally to noise or fluctuations effects. 

It is easy to see that in the case that the environment
 is modelled by a set of harmonic oscillators, the noise (Eq. (\ref{noise})) and 
 dissipation (Eq.(\ref{dissipation})) kernels become
\begin{equation}
\nu_{a}(t) = \frac{1}{2} \sum_n \lambda_{a,n}^2 \langle\left\{q_n(t), q_n(0)\right\}\rangle , \label{noiseqbm}
\end{equation}
\begin{equation}
\eta_{a}(t) =   \frac{1}{2} \sum_n \lambda_{a,n}^2 \langle\left[q_n(t), q_n(0)\right]\rangle , 
\label{dissipationqbm}
\end{equation} where $q_n$ are the position operators for the environmental degrees of freedom.

The noise correlations can be defined by their spectral density $J_a(\omega) = 1/(2\pi)
\int dt e^{i \omega t} 
\langle {\hat {\delta\omega_a}}(0) {\hat {\delta\omega_a}}(-s)\rangle_{\cal E}$ with 
$a = 0, 1,2$. If we assume the environment is composed by an infinite set of harmonic oscillators, it 
is useful to use the relation
\begin{equation}
\sum_n^N \frac{\lambda_n^2}{2 m_n \omega_n} f(\omega_n) = \int_0^\infty J(\omega) f(\omega) d\omega ,
\end{equation}
in order to express kernels in terms of integrals in frequency.
For example, using Eqs. (\ref{noiseqbm}) and (\ref{dissipationqbm}), the noise and dissipation 
kernels can be written as
\begin{eqnarray}
\nu_{a}(t) &=& \int_0^\infty J_{a}(\omega) \cos(\omega t) \coth(\frac{\beta \omega}{2}) d\omega,
\label{noiseqbm2}\\
\eta_{a}(t) &=&   \int_0^\infty J_{a}(\omega) \sin(\omega t)  d\omega,\label{dissipationqbm2}
\end{eqnarray}
where $\beta = 1/k_B T$ is the equilibrium temperature of the environment.

In this  model, we use  $J_a(\omega) = 
\gamma_a \omega \exp[-\omega/\Lambda]$ as the spectral density of the environment. 
This definition allows 
to calculate the noise and dissipation kernels from Eqs.(\ref{noiseqbm2}) and
(\ref{dissipationqbm2}) \cite{pra}.
In the definition of $J(\omega)$, 
$\Lambda$ is a physical ultraviolet cutoff, which represents the biggest 
frequency present in the environment. 

Starting with an arbitrary initial superposition state in the
Bloch sphere, $\vert \psi \rangle = \cos(\theta/2) \vert 0\rangle + \sin(\theta/2)\vert 1\rangle$, 
we numerically solve the master equation at different environmental temperatures. 
In the high temperature limit, one can expand the 
$\coth(\beta\omega/2)$ for small $\beta$, and obtain the noise kernel (the one that depends on 
temperature) as $\nu_a(t) =  \gamma_a k_B T \delta(t)$ (where, with the sub-index $a= 0$
we denote the longitudinal coupling and with the sub-index $a= 1,2$
transversal noise). In this limit, it is trivial to evaluate the diffusion terms 
in Eq.(\ref{coeficientes}), to obtain that 
$d_{xx} = \gamma_1 k_B T$, $d_{yy} = \gamma_2 k_B T$, $d_{zz} = \gamma_0 k_B T$, and all 
$f_{ij} = 0$ (there is no anomalous diffusion terms). 
In the opposite case, when $T \rightarrow0$, the diffusion kernel yields 
$\nu_a(t) = \gamma_a (t \Lambda \sin(\Lambda t) + \cos(\Lambda t) - 1)/t^2$. With this 
kernel all the diffusion coefficients in Eq.(\ref{coeficientes})  
can be obtained for a zero-T environment. It is important to remark that they are all  
different from zero and contribute to 
the master equation.  
We do not present here the explicit expression of them since 
their form is not relevant. The 
dissipative coefficients for the Ohmic environment, can be all calculated from the 
dissipation kernel Eq.(\ref{dissipationqbm2}). Thus, these  
kernels are given by $\eta_a(t) = \gamma_a \delta'(t)$, which are independent on 
temperature. Therefore, dissipation coefficients (in Eq.(\ref{coeficientes})) 
are $\gamma_{xy} = 2 \Delta \gamma_1$, 
$\gamma_{xz}= - 2 \gamma_1 \Omega_y$, $\gamma_{zx} = 2 \gamma_0 \Omega_y$, $\gamma_{zy} = 
- 2 \Omega_x \gamma_0$, $\gamma_{yx}= - 2 \gamma_2 \Delta$, 
and $\gamma_{yz} = 2 \gamma_2 \Omega_x$.

\begin{figure}[!ht]
\includegraphics[width=7.5cm]{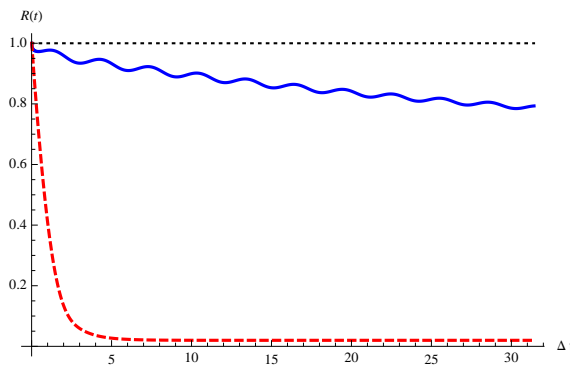}
\caption{(Color online) Temporal evolution of the absolute value of the Bloch vector for different environments coupled to the superconducting qubit. The black dotted line is the unitary evolution, 
i.e. no noise present in the evolution. The red dashed line represents an Ohmic environment
in the high temperature limit while the blue solid line represents an Ohmic environment at zero
temperature. It is easy to note that the state vector of the system is more affected by the
influence of the high temperature environment. However, we must note that the initially pure state
loses purity even at zero temperature for the same values of $\gamma_a$. We have considered the effect of longitudinal and transverse noise simultaneously. Parameters used: $\Omega_{\rm R}=0.1 \Delta,~\gamma_0=0.002=\gamma_1,~\Lambda=100 \Delta, ~T = 100 \Delta$.
We have set $\varphi_{\rm R}=0$ and $\gamma_2=0$ for simplicity. The initial state is given by $\theta = 2/3 \pi$ .}
\label{Fig1}
\end{figure}

\begin{figure}[!ht]
\includegraphics[width=7.5cm]{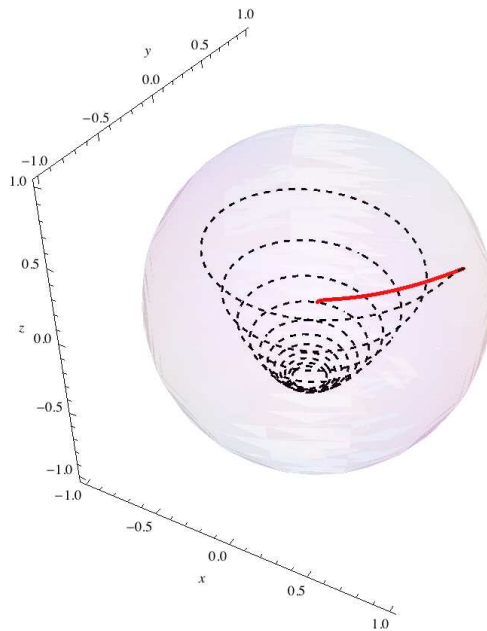}
\caption{(Color online). We present the trajectories of the qubit in the Bloch sphere for two periods of time. The black dashed line is the evolution for a qubit in the presence
of an Ohmic environment at zero temperature. The red solid line is for an Ohmic environment at high temperature. 
We have considered the effect of longitudinal and transverse noise simultaneously.
Parameters used: $\Omega_{\rm R}=0.1 \Delta,~\gamma_0=0.002=\gamma_1,~\Lambda=100 \Delta, ~T = 100\Delta$.
We have set $\varphi_{\rm R}=0$ and $\gamma_2=0$ for simplicity. The initial state is given by $\theta = 2/3 \pi$.}
\label{Fignuevaohmica}
\end{figure}

In Fig.\ref{Fig1} we present the absolute value of the Bloch vector of the state system $\bf{R}$ 
as a function of time for more than one period $\tau= 2 \pi /\tilde{\Omega}$, with $\tilde{\Omega}=\Delta^2/\sqrt{\Delta^2+\Omega_R^2}$.
Qualitatively, decoherence can be thought as the deviation of probabilities 
measurements from the ideal intended outcome. Therefore, 
decoherence can be understood as fluctuations in the Bloch vector ${\bf R}$ induced by noise. 
Since decoherence rate depends on the state of  the qubit, we will use as a measure of decoherence the change in 
time of the absolute value of ${\bf R}$, starting from $\vert {\bf R}\vert = 1$ for the initial 
 pure state, and decreasing as long as the quantum state losses purity. In Fig.\ref{Fig1}, the black dotted
 line is the unitary evolution (as expected ${\mathbf{R}}=1$ for all times), while
 the red dashed line is the evolution
 of the Bloch vector of a qubit evolving under a high temperature Ohmic environment. 
 This kind of environment is very destructive and the state vector is soon removed from the
 surface of the sphere (where purity states lie). The blue solid line
 represents the behaviour of
 the Bloch vector when the qubit is evolving under the influence of a zero temperature Ohmic environment. It is
 easy to note that the state losses purity even at zero temperature, 
 though the influence of the environment is not as drastic as when
 the temperature is high. We have considered the effect 
of the longitudinal and transverse noise simultaneously. These facts can also be seen in Fig.\ref{Fignuevaohmica}, where we plot the trajectories
 in the Bloch sphere. It is easy to see, that the Ohmic environment at high temperature 
 (red solid line) is
 very drastic, removing the state from the Bloch sphere surface in a very short timescale. 
 The qubit under the presence of an Ohmic environment at zero temperature (black dashed line) looses purity 
as well, but the timescale at which the quantum coherences are removed takes longer.

When we deal with a master equation of the form of Eq.(\ref{mastereq}), it is known that all real terms contribute
 to diffusion effects while imaginary ones to renormalisation and dissipation \cite{pla}. As decoherence is the 
 dynamical suppression of the quantum coherences, the off-diagonal elements of the reduced density matrix are also
 a good measure of how the environment affects the dynamics of the qubit.  For a general process (non-purely 
 dephasing process), the
 reduced density matrix can be represented as
 \begin{eqnarray}
 \rho_{\rm r}(t) &=&\bigg(\begin{array}{cc}
a_{00}(t)  & a_{01} {\cal F}(t)\\
a_{10} {\cal F}(t)^* &  a_{11}(t)
\end{array}\bigg), \nonumber
 \end{eqnarray}
 where ${\cal F}(t)$ can be related to the decoherence factor (which is responsible for the exponential decay of the
 coherences). In the case of  a longitudinal coupling only, the populations remain constant and ${\cal F}(t)$ is the decoherence 
 factor defined as $e^{-\int d_{zz}(t') dt'}$, where $d_{zz}(t)$ is the diffusion coefficient. 
 Independently of its formal expression, we know
 that ${\cal F}(t)$ must be a decaying function by which after some time bigger than the decoherence time $t > t_D$,
 the quantum coherences of the density matrix can be neglected. 
In the case of a more general coupling such as the one studied here, 
the decoherence factor is composed by the diffusion coefficient  ($d_{zz}(t)$, i.e. in this case) 
and some other anomalous diffusion and dissipation
coefficients
such as $f_{zx}(t)$ and $\gamma_{zy}(t)$, that affect the off-diagonal terms with no defined sign.

Due to the fact
that we are considering a general qubit (such as $H_{\rm eff}$ in Eq. (\ref{Hqubitdiag})) and a general coupling ($H_{\rm{int}}$ 
in Eq. (\ref{Hint})), there are some more coefficients 
that contribute to dissipation process such as $d_{xx}(t)$,
$f_{xz}(t)$, $f_{xy}(t)$, $f_{zy}(t)$ and $\gamma_{xy}(t)$  \cite{Benenti}. 
These complete set of coefficients provide a very complicated dynamics that
only can only be solved numerically. 

In order to study the decoherence process suffered by the qubit we propose also the study of the
off-diagonal terms as well as the absolute value of the Bloch vector. The difference between both quantities is that
while the latter one provides information about the purity of the state, the former gives information about the
decoherence timescale ($t_D$), i.e. the time after which the coherences of the reduced density matrix can be 
neglected. Therefore, we shall look at the behaviour of the quantum coherences, i.e the off-diagonal terms of the reduced density
matrix $\rho_{r_\textrm{off}}(t)$ to study the influence 
 of the environment on the system's evolution and derive, when possible, a decoherence timescale.
\begin{figure}[!ht]
\includegraphics[width=7.5cm]{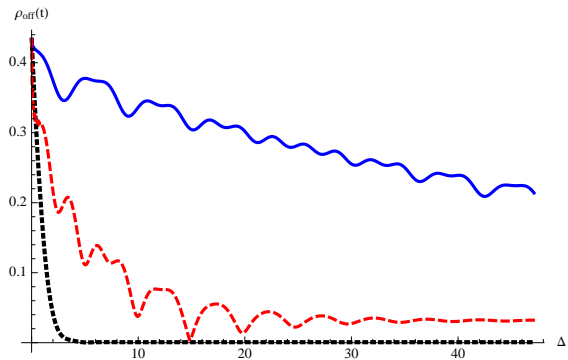}
\caption{(Color online) Temporal evolution of the absolute value of the quantum coherence (denoted
as $\rho_{r_\textrm{off}}(t)$),
namely the off-diagonal terms  of the qubit system ($\rho_{01}(t)$). 
 The black dotted line is the evolution of the coherences in the case the qubit is 
coupled to a high-T environment ($\gamma_0=\gamma_1=0.02$ and $T/\Delta = 100$). 
In this case, the 
system losses coherence quickly and with no possibility of re-coherence. The 
red dashed line, shows the evolution of the coherence in the limit of zero 
environmental temperature (with cutoff in frequencies $\Lambda = 100 \Delta$ and $\gamma_0=\gamma_1=0.02$). 
Finally, the blue solid oscillatory line represents the evolution of the coherence for a smaller
value of $\gamma_a$ at zero temperature ($\gamma_0=\gamma_1=0.002$). We have considered the effect of 
longitudinal and transverse noise simultaneously. It is easy to see that  the system also losses coherence 
when the environment is at zero temperature. The final value of the off-diagonal term differs from the one in 
the high-T case, due to the presence of diffusion and other dissipative coefficients in the master 
equation. We have considered  $\Omega_{\rm R}= 0.1 \Delta$, $\gamma_2=0$, $\varphi_{\rm R }=0$, and 
the initial state is given by $\theta = 2/3 \pi$.}
\label{Fig2}
\end{figure}
 
In Fig.\ref{Fig2} we present the evolution in time of the off-diagonal terms of the qubit's
reduced density matrix ($\rho_{r_\textrm{off}}(t)$) as a function of dimensionless time ($\Delta t$), 
in the case of high and zero temperature and for different dissipation constants
in the weak coupling limit. 
The black dotted line is the solution of the master equation in the 
limit of high temperature (for dimensionless parameter $T/\Delta = 100$). As  
expected, off-diagonal terms in the reduced density matrix decay quickly to 
their minimum value,  reaching a steady state of minimum coherence. A relevant 
result is the one obtained in the limit of $T =  0$ environmental temperature represented by
the blue solid line $\gamma_a=0.002$ and the red dashed one $\gamma_a= 0.02$. 
In this cases, we can see that, the 
coherences in the system decay (more slowly than in the case of high temperature) 
with time, reaching an asymptotic value of minimum coherence, at a timescale different from 
the one corresponding to the  high-T limit. The smaller the value of the coupling, the longer
it takes the system to reach the asymptotic value. This is mainly due to  
the presence, in the master equation for T = 0, of  anomalous diffusion coefficients, 
which are absent in the case of high-T. Nevertheless, we  show 
that fluctuations at zero-T also induce decoherence in 
the solid-state qubit, with a lower efficiency than in the thermal 
case, but strong enough to destroy the unitary evolution.  

In order to have a rough analytical estimation of decoherence times, we consider that the qubit is 
solely coupled in the longitudinal direction, and that there are no anomalous and dissipation 
terms in the master equation. This means that for the moment we neglect the effect of the tunnelling term 
(proportional to both transverse directions) in 
the main system Hamiltonian. Thus, we may follow the result given in Refs. \cite{pla,pra} for the purely dephasing model. 
There, decoherence time in the high temperature approximation can be estimated as $t_D \sim 2/(k_B T \gamma_0)$ \cite{makhlin}, 
which does not depend on the frequency cutoff $\Lambda$. Considering parameters used in Fig. \ref{Fig2}, 
one can estimate decoherence time to be $t_D \sim 1 \Delta$, in good agreement with the corresponding plots 
in Fig. \ref{Fig2}. 
For the Ohmic case at zero temperature, the decoherence scales as $t_D \geq 2/(\gamma_0 \Lambda \pi)$ for times
$\Lambda t \ge 1$. In this case, decoherence is delayed as $\gamma_0$ decreases. This is a very long bound for decoherence 
time, especially when the longitudinal coupling constant is very small. Indeed, this reflects the fact that the contribution of 
all the coefficients in the master equation (anomalous diffusion and dissipation coefficients) are important in the limit of zero 
temperature, as can be seen in Fig. \ref{Fig2}.\\

\begin{figure}[!ht]
\includegraphics[width=7.5cm]{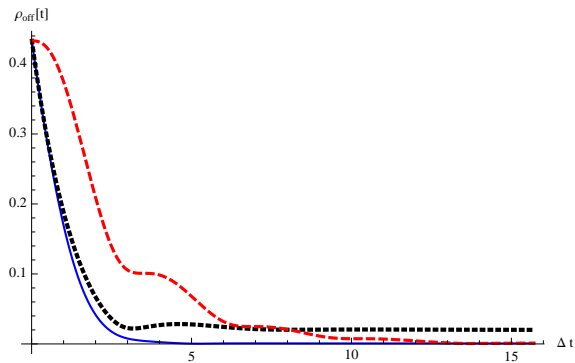}
\caption{(Color online) Temporal evolution of the coherence ($\rho_{r_\textrm{off}}(t)$)  
(its absolute value) of the qubit system coupled
to a high temperature Ohmic environment. The red dashed line represent the evolution
of the coherence when the qubit is coupled only in the transverse direction, namely through $\delta \omega_1$.
In this case, we use $\gamma_1=0.02$ and $\gamma_0=0$. The black dotted line is the evolution of the 
coherences in the case the qubit is 
coupled to a high-T environment only in the longitudinal direction, i.e. through $\delta \omega_0$.
This means $\gamma_1=0$ and for example, $\gamma_0=0.02$. Finally, the blue solid line (almost coincides with
the black dotted line) is the evolution
of the coherence when the qubit is equally coupled in both directions, i.e. $\gamma_0=0.02$ and 
$\gamma_1=0.02 $.
 We have considered  $\Omega_{\rm R}= 0.1 \Delta$, $\varphi_{\rm R}=0$, $\gamma_2=0$, and $T = 100\Delta$. 
The initial state is given by $\theta = \pi/3$.}
\label{Fig3}
\end{figure}

As we are considering a bidirectional coupling in our model, it is interesting to see if there is 
a direction in which noise becomes more important. As the value of $\gamma_0$ and $\gamma_1$
imply the coupling with the environment, we can turn off one of the couplings to study the effect
of noise in the longitudinal and transverse directions. Thus, we consider only one transverse direction to make
this analysis, by setting $\gamma_2=0$, since we consider the difference between longitudinal and transverse 
(either one) couplings. In Fig.\ref{Fig3} we follow the decay of the quantum coherences of the reduced density matrix
corresponding
to each of the cases considered: only longitudinal coupling, only transversal coupling and both couplings.
 The red dashed line represents the evolution
of the coherence when the qubit is coupled only in the transverse direction, namely through $\delta \omega_1$.
In this case, we use $\gamma_1=0.02$ and $\gamma_0=0$. The black dotted line is the evolution of the coherences in the case the qubit is 
coupled to a high-T environment only in the longitudinal direction, i.e. through $\delta \omega_0$.
This means $\gamma_1=0$ and for example, $\gamma_0=0.02$. Finally, the blue solid line (which falls together with
the black dotted line) is the evolution
of the coherence when the qubit is equally coupled in both directions, i.e. $\gamma_0=0.02$ and 
$\gamma_1=0.02$. It is easy to see that decoherence is mainly ruled by the longitudinal direction 
($t_D \sim 1 \Delta$) which
means that noise in the $\hat{z}$-direction affects more the unitary dynamics of the system than noise in $\hat{x}$
direction.

\begin{figure}[!ht]
\includegraphics[width=7.5cm]{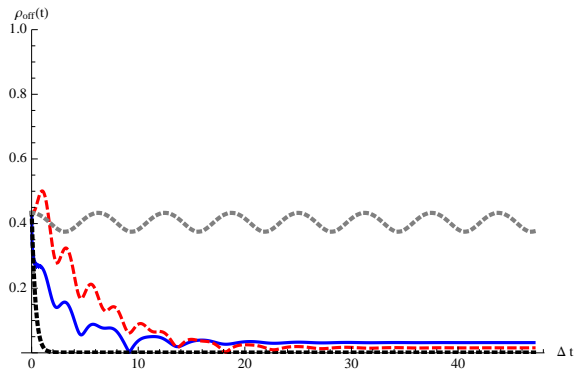}
\caption{(Color online) Temporal evolution of the absolute value of the coherence 
($\rho_{r_\textrm{off}}(t)$) of the qubit system coupled
to a zero-T Ohmic environment.
 The red dashed line is the evolution of the coherences in the case the qubit is 
coupled  only in the transverse direction, i.e. through $\delta \omega_1$.
This means $\gamma_0=0$ and for example, $\gamma_1=0.02$. The black dotted line represents the evolution
of the coherence when the qubit is equally coupled in both directions, i.e. $\gamma_0=0.02$ and 
$\gamma_1=0.02$. Finally, the blue solid line is the evolution
of the coherence when the qubit is coupled only in the longitudinal direction, namely through $\delta \omega_0$.
In this case, we use $\gamma_0=0.02$ and $\gamma_1=0$.
We have also included in a grey dotted line the oscillatory
unitary behaviour of the coherence.
 We have considered  $\Omega_{\rm R}= 0.1 \Delta$, $\varphi_{\rm R}=0$, $\gamma_2=0$, and the initial state is given by $\theta = \pi/3$.}
\label{Fig3bis2}
\end{figure}

We can also study the leading behaviour (related to the coupling of the system) 
 for a zero-T Ohmic environment. In Fig.\ref{Fig3bis2} we show the behaviour of the quantum coherences
  ($\rho_{r_\textrm{off}}(t)$) for different couplings. The black dotted  line
is a bidirectional (longitudinal and transverse) coupling of same value $\gamma_a=0.02$. The red dashed line is for a transverse coupling, i.e.
$\gamma_1=0.02$ and $\gamma_0=0$. Finally, the blue solid line is only a longitudinal coupling $\gamma_0=0.02$
and $\gamma_1=0$. We can note the longer decoherence timescale in comparison to an Ohmic environment in the 
high temperature limit, see for example Fig.\ref{Fig3}. It is also important to remark that at zero temperature,
all noises are equally important. This is a quiet important observation.
By taking a thorough look,
at shorter times, we can see that the blue solid line decays faster, corresponding to a longitudinal coupling. 
However, the timescale at which the longitudinal coupling decays is comparable to the timescale of a 
transversal coupling (in magnitude order). However, the intrinsic dynamics of the system delays 
the reaching of the
asymptotic limit, since we are studying the weak coupling limit.

%%%%%%%%%%%%%%%%%%%%%%%%%%%%%%%%%%%%%%%%%%%%%%%%%%%%%%%%%%%%%%%%%%%%%%%%%%%%%
%%%%%%%%%%%%%%%%%%%%%%%%%%%%%%%%%%%%%%%%%%%%%%%%%%%%%%%%%%%%%%%%%%%%%%%%%%%%%%%

\section{1/f noise} 
\label{1f}

Much effort has been spent recently to understand how noise at low frequencies 
affects the dynamics of superconducting qubits, both from a theoretical
and  experimental point of view. In solid-state systems decoherence is potentially strong due to numerous microscopic 
modes. Noise is dominated by material-dependent sources, such as background-charge fluctuations or 
variations of magnetic fields and critical currents, with given power spectrum, often known as $1/f$. This noise is 
difficult to suppress and, since the dephasing is generally dominated by the low-frequency noise, it is
particularly destructive.

The $1/f$ noise is frequently modelled by an ensemble of two-level systems or fluctuators and 
describes both Gaussian or non-Gaussian effects \cite{clarke,bergli}. 
Then, the noise is described as coming from $N$ uncorrelated fluctuators, that we call here 
$ {\hat {\delta\omega_N}} = \sum_i^N \chi_i(t)$, 
where $\chi_i(t)$ is a random telegraph process. The variable $\chi_i(t)$ takes the 
values $-\xi_i$ or $\xi_i$. Thus, 
$\chi_i(t)^2 = \xi_i^2 =$ const. 
By assuming a random process, there is no dissipation contribution.
In order to obtain the diffusion coefficients of the master equation, we
 need to evaluate the noise correlation functions from 
Eq.(\ref{noise}), for each of the interaction 
terms -the longitudinal and the transversal-, characterised 
by the subindex 0 and 1, respectively (we not consider the coupling in the 
${\hat y}$-direction in this Section for simplicity). We refer to these as 
\begin{equation}
\langle  \delta\hat\omega_{N,a}(t) 
\delta\hat\omega_{N,a} (s)\rangle =\sum_{i= 1}^N \xi_{i,a}^2 e^{-2 \zeta_{i,a} \vert t - s\vert},
\end{equation} where index $a = 0, 1$, indicates longitudinal and transversal couplings 
between the fluctuator and the qubit. 

Following Ref.\cite{nesterov}, we define the effective 
random telegraph process for $N\gg 1$, as $ {\hat {\delta\omega_a}} (t) 
= {\mbox lim}_{N\rightarrow \infty}  {\hat {\delta\omega_{N,a}}} (t)$, 
considering a continuous distribution of amplitudes ($\xi$) and switching rates ($\zeta$).
Assuming that for an individual fluctuator, the correlation relations are given by 
$
\langle \chi_i(t)\rangle =0$ and $\langle \chi_{i,a}(t)\chi_{j,a}(s)\rangle = 
\frac{\sigma_a^2}{N}\delta_{ij} e^{- 2 \zeta_a \vert t - s\vert}$; 
where $\sigma_a^2 = {\text{lim}}_{N\rightarrow\infty} N\xi_a^2$.  
For $N\rightarrow \infty$, the effective random process becomes a 
Gaussian Markovian process with an 
exponential correlation function. 
Finally, we consider that the noise correlation is defined by
\begin{equation}\langle  {\hat {\delta\omega_a}}(t)  
{\hat {\delta\omega_a}} (s)\rangle = {\sigma_a}^2 e^{-2 \zeta_a 
\vert t - s\vert}. \label{1fcorrelation}
\end{equation}
 
By using the noise correlation functions of Eq.(\ref{1fcorrelation}), 
we compute the diffusion coefficients Eq.(\ref{coeficientes}) of the master equation and 
solve it numerically to obtain the qubit dynamics.
We shall study the absolute value of the Bloch vector and the decay of
off-diagonal terms  to study the decoherence process suffered by the qubit, 
as we have done for the Ohmic environment.

\begin{figure}[!ht]
\includegraphics[width=7.5cm]{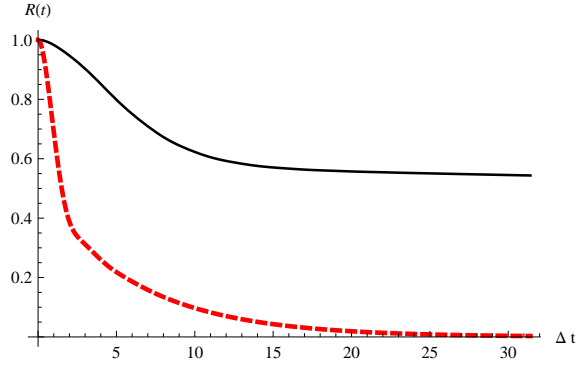}
\caption{(Color online) Time evolution of the absolute value of  Bloch vector for different values of the 
model parameters. The black solid line 
is the evolution for $\sigma_a=0.1\Delta$ and $\zeta_a=0.2 \Delta$. Red dashed line $\sigma_a= 0.5 \Delta$ and $\zeta_0=0.75 \Delta$. 
We have considered the effect of longitudinal and transverse noise simultaneously. 
We have set 
$\varphi_{\rm R}=0$ for simplicity, and the initial state is given by $\theta = \pi/3$.}
\label{Fignueva1f}
\end{figure}
In Fig.\ref{Fignueva1f} we present the temporal evolution of the absolute value of Bloch vector
$\bf{R}$ while
the qubit is evolving under the presence of $1/f$ noise.
We consider $\sigma_a  = 0.1 \Delta$ ($\zeta_a=0.2 \Delta$) for the black solid line 
and  $\sigma_a = 0.5 \Delta$ ($\zeta_a=0.75 \Delta$) for the red dashed curve ($a=0,1$). 
We can see that as the value of $\sigma_a$ becomes bigger, the sooner purity is lost.
This is so because $\sigma_a$ represents the coupling with the system ($a = 0$ in the longitudinal coupling, 
and $a= 1$ in the transverse case). Similar 
 to the case of an Ohmic environment at high-T, the 1/f noise is very 
efficient in inducing decoherence on the system.
The choice of parameters has been done to assure that there are no memory effects in the
evolution, namely $\zeta > \sigma$ in our weak coupled model. 
In both cases, we see that purity is a 
monotonic decaying function. 
\begin{figure}[!ht]
\includegraphics[width=7cm]{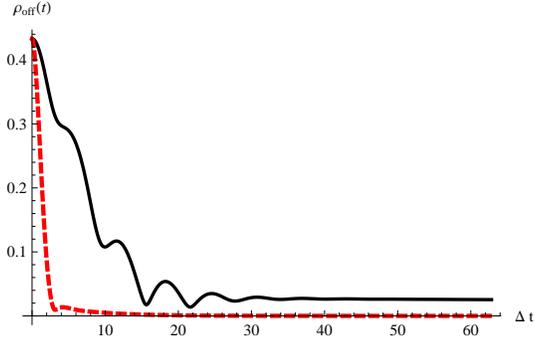}
\caption{(Color online) Decay of the quantum coherences  ($\rho_{r_{\text{off}}}(t)$), i.e the off diagonal terms
of the reduced density matrix $\rho_{r_{01}}(t)$
as function of time. The lines correspond to the same values of Fig.\ref{Fignueva1f}. We have considered the effect 
of longitudinal and transverse noise simultaneously. The initial state is given by $\theta = \pi/3$.}
\label{Fignueva1fbis}
\end{figure}

In order to have a rough estimation of the decoherence time, we consider only the longitudinal coupling
and no anomalous diffusion terms (i.e. dephasing process). Then, assuming 
that $\zeta_a t \le 1$, it is possible to show that $t_D \sim \sqrt{2}/\sigma_a$, independent of the 
switching rate. This fact can be clearly seen in Fig. \ref{Fignueva1fbis}, 
we have plotted the behaviour of the quantum coherences
of the reduced density matrix as function of time. It is worthy to note these decoherence timescales
are the temporal scale in which coherences abruptly decay from the pure-case value. This 
estimation sets a bound on the decoherence time. 
The asymptotic value is reached in a longer time. With the parameters used in Fig. \ref{Fignueva1f}, it is possible to check that decoherence time scales as 
$\Delta t_D \sim 3$ (red dashed line in Fig.\ref{Fignueva1fbis}), and $\Delta t_D \sim 14$ (see black solid line in 
Fig. \ref{Fignueva1fbis}). On the 
contrary, when $\zeta_a t \gg 1$, 
decoherence time scales as $t_D \sim 1/\zeta_a$, almost independently of the value of $\sigma_a$.
\begin{figure}[!ht]
\includegraphics[width=7.5cm]{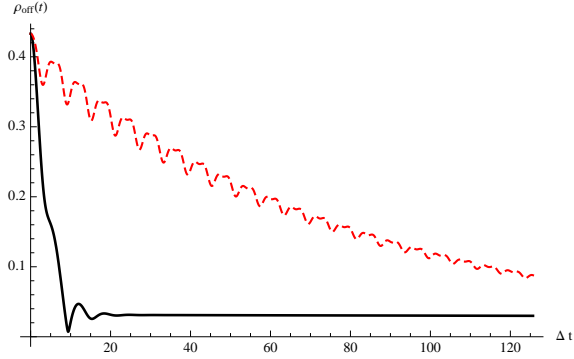}
\caption{(Color online) Time evolution of the absolute value of  Bloch vector in the Bloch sphere for different values of the 
model parameters. The red dashed line
is the evolution for $\sigma_1=0.2\Delta$, $\zeta_0=0.5\Delta$, $\sigma_0=0$. The black solid line is
for $\sigma_0=0.2\Delta$, $\zeta_1=0.5\Delta$, $\sigma_1=0$. We have set 
$\varphi_{\rm R}=0$ for simplicity. The initial state is given by $\theta = \pi/3$.}
\label{Fignueva1}
\end{figure}

Finally, we consider the effect of noise in both directions as we have done for the Ohmic environments. 
It is important to note that in the case of $1/f$ noise, 
the coupling constant is included in parameter $\sigma_a$ of the model.
Here, we will study the behaviour of the coherence $\rho_{r_{\text{off}}}(t)$ 
to infer how decoherence is induced in each case.
In the following figures we present the behaviour of the quantum state for different coupling situations.
\begin{figure}[!ht]\includegraphics[width=7.5cm]{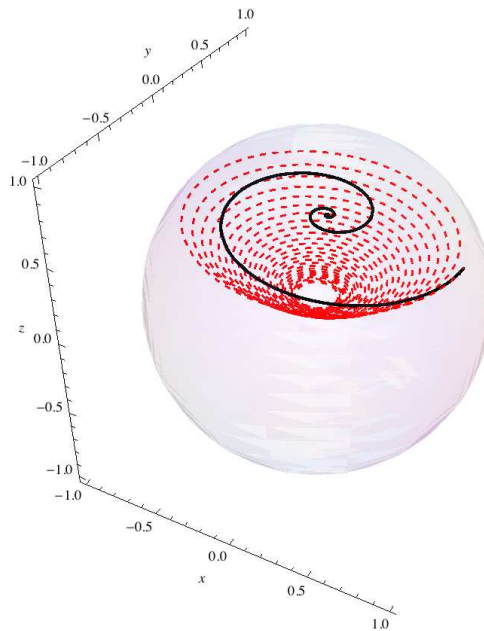}
\caption{(Color online) Trajectories in the Bloch sphere for the values of the model parameters
of Fig.\ref{Fignueva1}. The red dashed line
is the evolution for $\sigma_1=0.2\Delta$, $\zeta_0=0.5\Delta$, $\sigma_0=0$. The black solid line is
for $\sigma_0=0.2\Delta$, $\zeta_1=0.5\Delta$, $\sigma_1=0$. We have set 
$\varphi_{\rm R}=0$ for simplicity. The initial state is given by $\theta = \pi/3$}
\label{Fignueva2}
\end{figure}
In Fig.\ref{Fignueva1} we present the temporal evolution of the absolute value of Bloch vector
$\bf{R}$ while 
the qubit is evolving under the presence of $1/f$ noise and in Fig.\ref{Fignueva2} 
the trajectories in the Bloch sphere for the same model parameters. 
The red dashed line
is the evolution for $\sigma_1=0.2\Delta$, $\zeta_0=0.5\Delta$, $\sigma_0=0$. The black solid line is
for $\sigma_0=0.2\Delta$, $\zeta_1=0.5\Delta$, $\sigma_1=0$. 
This means that the black solid line represents a qubit coupled to the environment in a 
longitudinal direction only  while the red dashed
is the qubit coupled to the environment in the transverse
direction only. It is easy to see in Fig.\ref{Fignueva2} that the longitudinal coupling removes the state
from the surface of the sphere rapidly by loosing purity faster (Fig.\ref{Fignueva1}).  In the case of the
transverse noise (red color), the effect of noise on the system may be neglected for very short times, by noting
that the trajectory remains very similar to the unitary one in that timescale.
The black solid line shows that having a coupling in the $\hat{z}$ direction removes the system for
the Bloch sphere faster than  having a coupling only in the $\hat{x}$ direction. 
This means that the purely dephasing process is the leading process when having both couplings together.

%************************************************************************************
%************************************************************************************
%************************************************************************************
%************************************************************************************
\section{Final Remarks}
\label{conclusions}
The interaction of a solid-state qubit with environmental degrees of freedom strongly 
affects the qubit dynamics, and leads to decoherence. 
In quantum information processing with solid-state qubits, decoherence significantly 
limits the performance of such devices. These degrees of freedom 
appear as noise induced in the parameters entering the qubit Hamiltonian and also as 
noise in the control currents. These noise sources produce decoherence 
in the qubit, with noise, mainly, at microwave frequencies affecting the relative population between 
the ground and excited state, and noise or low-frequency fluctuations affecting the phase of the 
qubit. It is important to study the physical origins of decoherence by 
means of noise spectral densities and noise statistics. 
Therefore, it is necessary to fully understand the mechanisms that lead to decoherence.

We have derived a perturbative master equation for a superconducting qubit coupled to external sources of noise,
including the combined effect of noise in the longitudinal and 
transversal directions. We have considered different types of noise by defining their correlation function in time.
Decoherence can be understood as fluctuations in the Bloch vector ${\bf R}$ induced by noise. 
Since decoherence rate depends on the state of 
 the qubit, we have represented decoherence by the change of $\vert{\bf R}\vert$ in time, 
starting from $\vert {\bf R}\vert = 1$ for the initial 
 pure state, and decreasing as long as the quantum state losses purity. 

For an Ohmic environment, we have considered thermal effects. We have solved the master equation and presented
the dynamics of the superconducting qubit in the presence of a high temperature environment and a zero temperature
one. In both cases, we have computed the corresponding diffusion and dissipation coefficients, solved the master
equation numerically and derived
some analytical rough estimations of the decoherence time when possible. 
As expected, 
an environment at high temperature is an effective coherence destructor and a pure state vector
is soon removed from the surface of the Bloch sphere. In addition, we have shown that decoherence
is still induced in the qubit when the environment is at zero temperature. This process is less drastic and
takes longer times compared to the high temperature limit. However, it is important to remark that the decoherence
process still takes place, this time induced by the vacuum fluctuations of the environment. This fact has been shown 
for example in the
loss of purity of the state vector and in the decay of the off-diagonal terms of the reduced density matrix, namely 
coherences.
This result is in contrast to some recent publications \cite{pekola}, where the
effect of the environment at zero temperature is neglected.
We have also focused on the effect of longitudinal and transversal noise. As expected, when the qubit is coupled 
to both directions, namely longitudinal and transverse, the influence of the environment is bigger as observed in the
destruction of the coherences. However, it is important to note that having a transverse coupling only, does not imply 
a decoherence process as important as the one induced by the system when the coupling is longitudinal and the environment
is at high temperature. This result
is novel and should help in future qubits designs or 
experimental setups. Another important fact to take into account is that in the case of a zero-T environment,
all noises are equally important at short times. This means that both noises affect the dynamics of the superconducting 
qubit in the same decoherence timescale as has been shown.
The difference between couplings has not been studied before and this result is important when experimental setups
are considered to be done in zero-T conditions.

We have also discussed the role of low-frequency of decoherence on quantum bits, namely a noise $1/f$, 
modelled herein by an ensemble of two-level fluctuators. We have presented this analysis in the framework of the  
master equation approach. From the definition of the noise correlation function of the environment, we
have computed the diffusion coefficients and solved numerically the dynamics of the qubit. We have studied
how this type of noise affects the coherences of the reduced density matrix and how the state
vector is removed from the surface of the Bloch sphere. We have seen that this noise can be very destructive,
depending on the value of the free parameter $\zeta$. We have provided some rough analytical estimations of
the decoherence timescale that agree with the numerical solutions presented here. 
As for the effect of longitudinal and transversal noise, when the coupling is bidirectional the effect of noise is
bigger on the coherences of the qubit.

The analysis of the decoherence timescales may provide additional information
about the statistical properties of the noise. 
The comprehension of the
decoherence and dissipative processes, origin and causes,
should allow their further suppression in future qubits designs or 
experimental setups.

%************************************************************************************
%************************************************************************************
%************************************************************************************
%************************************************************************************
%\section{Appendix}

\acknowledgments

This work is supported by CONICET, UBA, and ANPCyT, Argentina.

\appendix
\section{}\label{A}
The functions $X_{a}, Y_{a}$, and $Z_{a}$ appearing in Eq.(\ref{coeficientes}) 
are derived by obtaining the temporal dependence of the 
Pauli operators $\sigma_i$ in the Heisenberg representing through the differential equations

 \begin{equation}
 \frac{d\sigma_k(t)}{dt} = i \left[H_q, \sigma_k(t)\right],
 \end{equation}
 with $k = x, y, z$ and $H_{\rm eff}$ as in Eq.(\ref{Hqubitdiag}). The solution can be expressed as a 
linear combination of the Pauli matrices  (in the Schr\"odinger
 representation) as 
$ \sigma_z^{a} = X_{a}(t) \sigma_x + Y_{a}(t) \sigma_y  + Z_{a}(t) \sigma_z$. 
 The explicit solution is given by
 
\begin{widetext}
\begin{eqnarray}
 X_1(t) &=& \frac{\Omega_x^2 + \left(\Delta^2  + \Omega_y^2\right) \cos(2 t \sqrt{\Omega_{\rm R}^2 + \Delta^2})}{\Omega_{\rm R}^2 + \Delta^2},\nonumber \\
 Y_1(t) &=& \frac{\Omega_x \Omega_y \left(1 - \cos(2 t \sqrt{\Omega_{\rm R}^2 + \Delta^2})\right) +  \Delta \sqrt{\Omega_{\rm R}^2 + \Delta^2} \sin(2 t \sqrt{\Omega_{\rm R}^2 + \Delta^2})}{\Omega_{\rm R}^2 + \Delta^2},\nonumber \\
Z_1(t) &=&  \frac{\Omega_x \Delta \left(1 - \cos(2 t \sqrt{\Omega_{\rm R}^2 + \Delta^2})\right) -  \Omega_y \sqrt{\Omega_{\rm R}^2 + \Delta^2} \sin(2 t \sqrt{\Omega_{\rm R}^2 + \Delta^2})}{\Omega_{\rm R}^2 + \Delta^2},\nonumber \\
X_2(t) &=& \frac{\Omega_x \Omega_y \left(1 - \cos(2 t \sqrt{\Omega_{\rm R}^2 + \Delta^2})\right) -  \Delta \sqrt{\Omega_{\rm R}^2 + \Delta^2} \sin(2 t \sqrt{\Omega_{\rm R}^2 + \Delta^2})}{\Omega_{\rm R}^2 + \Delta^2}, \nonumber \\
Y_2(t) &= & \frac{\Omega_y^2 + \left(\Delta^2  + \Omega_x^2\right) \cos(2 t \sqrt{\Omega_{\rm R}^2 + \Delta^2})}{\Omega_{\rm R}^2 + \Delta^2},\nonumber \\
Z_2(t) &= &  \frac{\Delta \Omega_y \left(1 - \cos(2 t \sqrt{\Omega_{\rm R}^2 + \Delta^2})\right) +  \Omega_x \sqrt{\Omega_{\rm R}^2 + \Delta^2} \sin(2 t \sqrt{\Omega_{\rm R}^2 + \Delta^2})}{\Omega_{\rm R}^2 + \Delta^2},\nonumber \\ 
X_0(t) &=& \frac{\Delta \Omega_x \left( 1 - \cos(2 t \sqrt{\Omega_R^2 + \Delta^2})\right) + \Omega_y \sqrt{\Omega_{\rm R}^2 + \Delta^2} \sin(2 t \sqrt{\Omega_R^2 + \Delta^2})}{\Omega_R^2 + \Delta^2},\nonumber \\
 Y_0(t) &=& \frac{\Delta \Omega_y \left( 1 - \cos(2 t \sqrt{\Omega_R^2 + \Delta^2})\right) - \Omega_x \sqrt{\Omega_{\rm R}^2 + \Delta^2} \sin(2 t \sqrt{\Omega_R^2 + \Delta^2})}{\Omega_R^2 + \Delta^2},\nonumber \\
 Z_0(t) &=& \frac{\Delta^2 + \Omega_{\rm R}^2 \cos(2 t \sqrt{\Omega^2 + \Delta^2})}{\Omega_{\rm R}^2 + \Delta^2}
 .\nonumber 
 \end{eqnarray} 
\end{widetext}
With these functions, we can evaluate all the coefficients in Eq.(\ref{coeficientes}), and obtain the desired temporal evolution 
for each of the considered types of environment, defined through the spectral density.

\end{document}